\documentstyle[aps,prd,preprint,epsf]{revtex}
\tighten
\newcommand{\ee}{e^{+}e^{-}}

\newcommand{\jp}{J/\psi}
\newcommand{\ec}{\eta_{c}}
\newcommand{\ecp}{\eta_{c}(2S)}
\newcommand{\psip}{\psi(2S)}

\newcommand{\pipi}{\pi^{+}\pi^{-}}
\newcommand{\kpi}{K^-\pi^{+}}
\newcommand{\ks}{K_{S}}

\newcommand{\Bpl}{B^{+}}
\newcommand{\kpl}{K^{+}}

\newcommand{\rt}{\rightarrow}

\newcommand{\etal}{\em et al.}
\title{
Observation of the  {\boldmath $\ecp$}  in exclusive
{\boldmath $B \rt K \ks\kpi$} decays}  

 \author{
  S.-K.~Choi$^{7}$,           
  S.~L.~Olsen$^{8}$,          
  K.~Abe$^{9}$,               
  K.~Abe$^{43}$,              
  R.~Abe$^{30}$,              
  T.~Abe$^{44}$,              
  I.~Adachi$^{9}$,            
  Byoung~Sup~Ahn$^{16}$,      
  H.~Aihara$^{45}$,           
  M.~Akatsu$^{23}$
% ...
  Z.~P.~Zhang$^{37}$,         
  V.~Zhilich$^{2}$,           
and
  D.~\v Zontar$^{50}$         
}
\author{(The Belle Collaboration)}
\author{~}
\address{
% ...
$^{2}${Budker Institute of Nuclear Physics, Novosibirsk}\\
% ...
$^{7}${Gyeongsang National University, Chinju}\\
$^{8}${University of Hawaii, Honolulu HI}\\
$^{9}${High Energy Accelerator Research Organization (KEK), Tsukuba}\\
% ...
$^{16}${Korea University, Seoul}\\
% ...
$^{23}${Nagoya University, Nagoya}\\
% ...
$^{30}${Niigata University, Niigata}\\
% ...
$^{37}${University of Science and Technology of China, Hefei}\\
% ...
$^{43}${Tohoku Gakuin University, Tagajo}\\
$^{44}${Tohoku University, Sendai}\\
$^{45}${University of Tokyo, Tokyo}\\
% ...
$^{50}${University of Tsukuba, Tsukuba}\\
% ...
}

\date{\today}
\begin{document}
\maketitle

\begin{abstract}

We report the observation of a narrow peak in the
$\ks\kpi$ invariant mass distribution in a sample of
exclusive $B \rt K \ks\kpi$ decays collected with the 
Belle detector at the KEKB asymmetric $\ee $ collider.  
The measured mass of the peak is  
$M = 3654 \pm 6 (stat) \pm 8 (syst) $~MeV/$c^2$ and 
we place a 90\% confidence level
upper limit on the width of $\Gamma < 55$~MeV/$c^2$. The
properties agree
with heavy-quark potential model expectations for the 
$\ecp$ meson, the $n=2$ singlet $S$ charmonium state.

\end{abstract}

\vspace*{0.5cm}
%

Major experimental issues for the charmed-quark
anticharmed-quark
($c\overline{c}$)  charmonium particle
system are the two $c\overline{c}$ states that are expected to 
be below open charm threshold but
are still not well established: the radially excited
$n=2$ singlet $S$ state, the
$\eta_c(2S)$ meson, and the $n=1$ singlet $P$ state, the 
$h_c(1P).$  The
observation of these states and the determination of
their masses would complete the below-threshold charmonium
particle spectrum and provide useful information about
the spin-spin part of the charmonium potential~\cite{Rosner}.

$B$ meson decays provide an excellent opportunity for
searching for the $\ecp$ and clarifying its properties.
They are a copious $\ec (1S)$ source:  
the decays $B \rt K \ec (1S)$  have been
observed by CLEO~\cite{Edwards}, BaBar~\cite{Aubert}, and
Belle~\cite{ffang} with relatively large
branching fractions:
${\cal B}(B\rt K \ec (1S) ) \simeq {\cal B}(B\rt K \jp) \simeq 1 \times
10^{-3}$.  (In the following, we use $\ec$ to denote the $\ec (1S)$.)
Moreover, in the case of the triplet charmonium $S$ states,
$B$ meson decays to the radially excited $\psip$ 
are nearly as common as those to
the $n=1$ $\jp$ radial ground state:
${\cal B}(B^+\rt K^+ \psip)/{\cal B}(B^+\rt K^+ \jp ) \sim
0.6$~\cite{PDG}.
Thus, it is reasonable to expect the decays $B\rt K \ecp $ to occur
at a rate comparable to that for $B\rt K \ec $.  Unlike the
$\jp$ and $\psip$, where hadronic decays proceed via
highly suppressed three-gluon intermediate states, the
$\ec $ and $\ecp$ decay via less-suppressed two-gluon processes.
As a result, intercharmonium transitions are not very
important and the hadronic decay branching fractions for the 
two states are expected to be similar~\cite{SFT}. Thus,
any final state that shows a strong $B\rt K\ec$ signal is
a promising channel for an $\ecp$ search. 

A simple application of heavy-quark potential models~\cite{QQ_potential}
predicts a $\psi(2S)$-$\ecp$ mass splitting that is smaller than
that for the ground-state $\jp$-$\ec$ splitting because of
the smaller value of the
wave function at zero $c\overline{c}$ separations and
the running of the QCD coupling strength 
$\alpha_s(M^2)$.  These models predict an $\ecp$ mass in
the range $3625 < M_{\ecp} < 3645 $~MeV/$c^2$.
Similar factors result in the expectation that the $\ecp$ total
width is somewhat smaller than that of the $\ec$.

The Crystal Ball group~\cite{XTAL}
reported an excess of $E_{\gamma}\simeq 91$~MeV
gamma rays from inclusive $\psip\rt\gamma X$ decays,
and interpreted this as possible
evidence for the $\eta_c(2S)$ with mass $3594 \pm 5$~MeV/$c^2$.
This result implies a $\psi(2S)$-$\ecp$ mass splitting that
is considerably larger than heavy-quark potential model
expectations.  The result has not 
been confirmed by other experiments~\cite{E760}.

In this letter we report a search for the $\ecp$ produced via
the processes  $\Bpl \rt \kpl \ecp $ and $B^0 \rt \ks \ecp$,
where $\ecp\rt\ks\kpi$~\cite{CC}.  We concentrate on this 
final state because it
is a strong decay channel for the $\ec $ (${\cal B}\simeq 1.8\%$),
has low combinatorial backgrounds, and, since the final
state contains all charged particles, is reconstructed with
good resolution.  Moreover, the process $\psip \rt \ks\kpi$ is
strongly suppressed, and the background in this channel from 
$B\rt K\psip$ decays is expected to be less than 0.1 events.

The search uses 
a 41.8~fb$^{-1}$ data sample collected with the Belle
detector~\cite{Belle} at the KEKB $e^+e^-$ collider~\cite{KEKB} 
operating at the $\Upsilon(4S)$ resonance ($\sqrt{s} = 10.58$~GeV). 
The data sample contains 44.8 million $B\overline{B}$ meson
pairs.

The Belle detector is a large-solid-angle magnetic
spectrometer that
consists of a three-layer silicon vertex detector,    
a 50-layer central drift chamber (CDC), an array of
aerogel threshold \v{C}erenkov counters (ACC),
a barrel-like arrangement of time-of-flight
scintillation counters (TOF), and an electromagnetic calorimeter
comprised of CsI(Tl) crystals  located inside
a superconducting solenoid coil that provides a 1.5~T
magnetic field.  An iron flux-return located outside of
the coil is instrumented to detect $K_L$ mesons and to identify
muons.  The detector
is described in detail elsewhere~\cite{Belle}.

We select events with
$\kpl\ks K^{\pm}\pi^{\mp}$ or
$\ks\ks K^{\pm}\pi^{\mp}$ combinations.
Here the charged kaon (pion) tracks are required
to originate from within $\delta r < 0.3$~cm
and $\vert \delta z \vert < 2.2$~cm of the  
run-by-run determined interaction point (IP)
in the transverse ($r\phi$) and beamline ($z$) directions, respectively.
In addition, they must be positively identified as kaons 
(pions) by the combined information from the ACC,
TOF and CDC $dE/dx$ measurement.  Candidate
$\ks\rt\pipi$ decays correspond to pairs of oppositely charged
tracks with invariant mass within 12~MeV/$c^2$ ($3\sigma$)
of $M_{K^0}$ that originate from a common vertex that
is displaced by more than 0.3~cm from the IP.  The
direction of the $\ks$ momentum vector is required
to be within 0.2~rad of the direction between
the IP and the position of the displaced  vertex.

Candidate $B$ mesons are reconstructed using
the energy difference 
$\Delta E\equiv E_B^{\rm cms} - E_{\rm beam}^{\rm cms}$
and the beam-energy constrained
mass $M_{\rm bc}\equiv\sqrt{(E_{\rm beam}^{\rm cms})^2-(p_B^{\rm
cms})^2}$,
where $E_{\rm beam}^{\rm cms}$ is the center of mass (cms) beam
energy,
and $E_B^{\rm cms}$ and $p_B^{\rm cms}$ are the cms energy and
momentum of the $B$ candidate.
The signal region is defined as 5.271 $< M_{\rm bc} <$ 5.287 GeV/$c^2$
and $|\Delta E |< $ 0.040 GeV, which correspond to $\pm 3\sigma$ from
the central values for both quantities.

In order to suppress background from the $\ee \rt q\overline{q}$ 
continuum ($q = u,~d,~s~\&~c$),  we form a likelihood
ratio  from two variables.   One is a Fisher
discriminant determined from five modified Fox-Wolfram
moments~\cite{SFW}, the cosine of the angle formed by
the thrust axis of the candidate $B\rt K \ks\kpi$ tracks
and that of the remaining tracks in the event, and
the sum of the absolute values of transverse momenta of 
particles relative to the $B$ candidate's thrust axis 
with angle larger than $60^{\circ}$ normalized by the sum of the total
momenta.  The coefficients of the Fisher discriminant are chosen to
optimize the separation between signal and continuum Monte Carlo (MC)
events~\cite{MC}.  The other is
the cosine of the angle
between the $B$ candidate flight direction and the beam axis in
the $\Upsilon(4S)$ rest frame ($\cos\theta_B$).
Normalized probability density functions (pdfs) formed
from the Fisher discriminants and the $\cos\theta_B$
distribution
are multiplied to form likelihood
functions for the signal (${\mathcal L}_{sig}$)
and continuum (${\mathcal L}_{cont}$) processes.
We select events with a likelihood ratio 
$LR\equiv {\mathcal L}_{sig}/({\mathcal L}_{sig}
+ {\mathcal L}_{cont})>0.6$, which was determined by
optimizing $S/\sqrt{S + B}$ ($S$ and $B$ are signal and background,
respectively) for Monte Carlo simulations of the process 
$B\rt K \ec $, where $\ec\rt\ks\kpi$.

We reduce potential backgrounds from
$B\rt D(D_s) X$ decays by rejecting $D$ and $D_s$ mesons
with the requirements
$\vert M_{K\pi}-M_{D} \vert > 10$~MeV/$c^2$ 
and $\vert M_{\ks\kpl}-M_{D_s} \vert > 10$~MeV/$c^2$.
The decay
$\eta_c(nS)\rt K^*(890) K$ is suppressed by an angular
momentum barrier; in order to reduce backgrounds from other
$B$ meson decay modes with minimal loss in signal,
we reject events with a $K^*$ candidate with
the requirement
$\vert M_{K\pi}-M_{K^*} \vert > 50$~MeV/$c^2$.

\begin{figure}[htb]
\centerline{\epsfysize 5.5 truein
\epsfbox{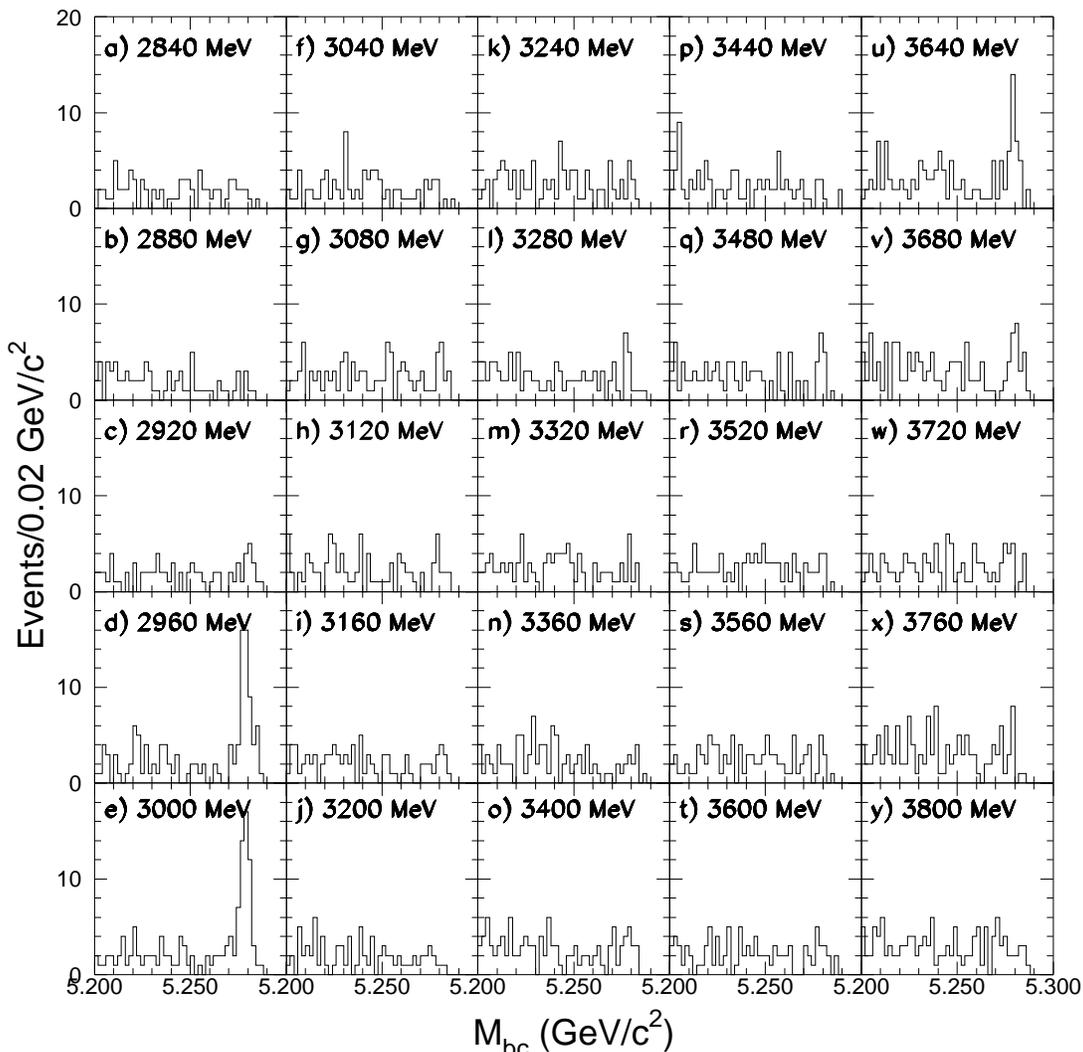  }}
\caption{\label{fig:mb_25box}
The $M_{\rm bc}$ projections
for 40 MeV/$c^2$ bins of
$M_{\ks K\pi}$, with central values
ranging from 2840 (a) to 3800~MeV/$c^2$ (y).
Only events with $\vert \Delta E \vert < 40$~MeV are
included; the charged and neutral $B$ decay modes are combined.}
\end{figure}

Figure~\ref{fig:mb_25box} shows 
the $M_{\rm bc}$ projections of events in the
$\vert \Delta E \vert \le 0.040$~GeV signal region
for twenty five $M_{\ks K\pi}$ mass bins,
each 40~MeV/$c^2$-wide and with
central values ranging from 2840 through 3800~MeV/$c^2$.
The mass bins of 
Figs.~\ref{fig:mb_25box}(d) 
and (e) straddle $M_{\ec}$ and clear peaks 
corresponding to $B\rt K\ec$, $\ec\rt\ks\kpi$ decays are apparent.  
Figures~\ref{fig:mb_25box}(u) and (v), which cover
a region near the expected mass of the $\ecp$, 
also show distinct $B$ meson signals.

\begin{figure}[htb]
\centerline{\epsfysize 2.0 truein
\epsfbox{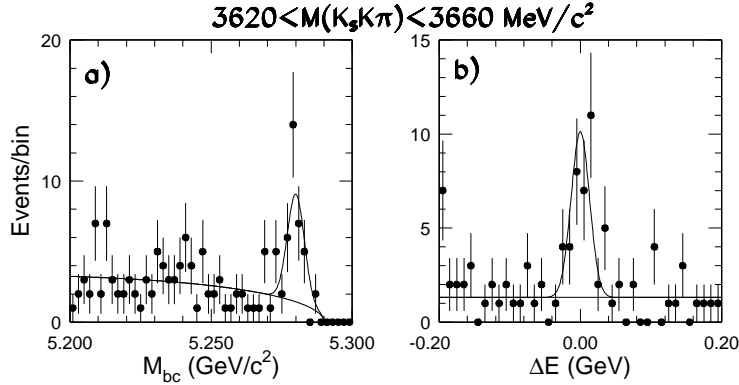  }}
\caption{\label{fig:de_mb_kskpi_2box}
The (a) $M_{\rm bc}$ and (b) $\Delta E$ projections
for the $M_{\ks K\pi} = 3640$~MeV/$c^2$ mass bin.
The curves are the results of the simultaneous fit described
in the text.}
\end{figure}

We perform simultaneous fits
to each of the $M_{\rm bc}$ distributions of Fig.~\ref{fig:mb_25box}
and the corresponding $\Delta E$ distributions for events 
in the $M_{\rm bc}$ signal region (not shown).  The fits use 
Gaussian functions with MC-determined widths to represent 
the signals; the areas of the $M_{\rm bc}$ and $\Delta E$ signal 
functions are constrained to be equal.  The $M_{\rm bc}$ background 
is modeled by a smooth function that behaves like phase-space near the
kinematic end point~\cite{ARGUS}; for the $\Delta E$ background, 
we use a second-order polynomial.  As an example, the results of the
fit to the $M_{\rm bc}$ and $\Delta E$ distributions
of the $M_{\ks K\pi} = 3640$~MeV/$c^2$ bin are shown in 
Figs.~\ref{fig:de_mb_kskpi_2box}(a) and (b), respectively.

\begin{figure}[htb]
\centerline{\epsfysize 2.0 truein
\epsfbox{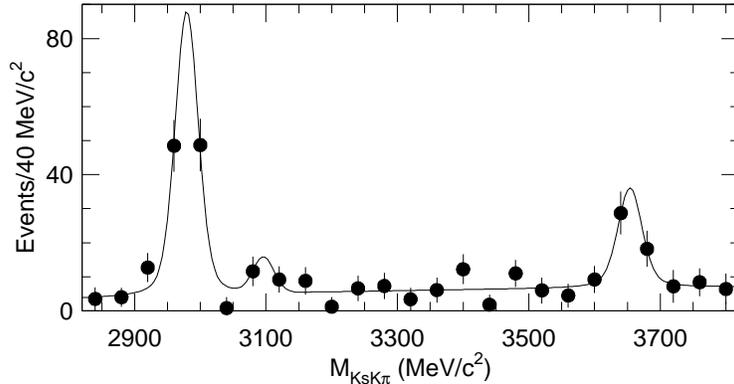}}
\caption{\label{fig:sliced_results} 
The distribution of signal events from
the simultaneous fits to $M_{\rm bc}$ and $\Delta E$ for each
$\ks K\pi$ mass bin.  The curve is the result of the
fit described in the text. }
\end{figure}

The signal yields extracted from the
simultaneous fits to the different $\ks\kpi$ mass bins
are plotted {\it vs.} $M_{\ks K\pi}$ in
Fig.~\ref{fig:sliced_results}, where, in addition 
to a prominent $\ec$  peak
and a hint of a $J/\psi$, a clear peak
at higher mass is evident.  We identify this as a
candidate for the $\ecp$. Between the peaks is a non-zero,
non-resonant contribution.  The curve in Fig.~\ref{fig:sliced_results}
is the result of a fit with simple Breit-Wigner functions 
that represent the $\ec$ and candidate $\ecp$, a Gaussian function with
mass and width fixed at the
$J/\psi$ values, and a second-order polynomial to represent the
non-resonant contribution.  These functions are convolved with a Gaussian
resolution function with a MC-determined width of
$\sigma=15$~MeV/$c^2$.

The fit values
for the event yields, masses and total widths of the $\ec$ and
the  $\ecp$ candidate state are listed with
their statistical errors in 
Table~\ref{table:fit_results}.  The fit value for the
$\ec$ mass is in good agreement with the world-average value
of $M_{\ec} = 2979.8 \pm 1.8$~MeV/$c^2$~\cite{PDG}; the 
value for the $\ec$ width is consistent, within its
rather large errors, both with the existing world average of
$\Gamma_{\ec}^{\rm tot} = 13.2^{+3.8}_{-3.2}$~MeV/$c^2$~\cite{PDG}
and the recent CLEO result of $26\pm 6$~MeV/$c^2$~\cite{Branden}.

The sum of the observed events in the three
mass bins in the signal region
(i.e., centered around $M(\ks K\pi)=3640$~MeV/$c^2$) 
is 56, while the integral of the second-order polynomial 
over the same interval
gives a non-resonant expectation of $21 \pm 2$ events.
The probability for this 
to fluctuate up to 56 events 
is $\sim 10^{-8}$, which corresponds to a signal
significance of more than 6$\sigma$.

The fitted mass of the candidate $\ecp$ is
substantially above the Crystal Ball mass value
and consistent, within errors, with the upper
end of potential model expectations. 
The systematic error on the mass
is evaluated by redoing the analysis 
using different likelihood ratio selection requirements,
50~MeV/$c^2$-wide bins, bins with central values
shifted by half a bin-width, and with different
values of the experimental resolution.   The maximum 
change in the fitted mass value is 8~MeV/$c^2$, which
is taken as the systematic error.  The limited
statistics and the 
resolution precludes a precise measurement of the
width.  However,  we can establish a 90\% confidence
level upper limit of $\Gamma <55$~MeV/$c^2$.

Monte Carlo simulations indicate that the
acceptance is very nearly constant for the
$\ks\kpi$ mass region covered by this measurement~\cite{accept}.
Thus, the ratio of product branching
fractions for the $\ec$ and $\ecp$ is just the ratio 
of event yields:
\begin{equation}
\frac{{\cal B}(B\rt K\ecp ){\cal B}(\ecp\rt\ks\kpi)} 
{{\cal B}(B\rt K\ec ){\cal B}(\ec \rt\ks\kpi)}
 = 0.38\pm 0.12 \pm 0.05,
\end{equation}
where the first error is statistical and the second
systematic.  The systematic error is determined
from changes in the ratio observed for different
binning, values of 
resolution, and functions used to
model the non-resonant contribution.

In summary, we observe a peak
in the $\ks\kpi$ mass from exclusive 
$B^+\rt K^+ \ks\kpi $ and
$B^0\rt \ks \ks\kpi $ 
decays with mass and width values
\begin{eqnarray*}
M       & = & 3654 \pm 6 \pm 8~{\rm MeV}/c^2 \\
\Gamma  & < & 55 {\rm ~MeV}/c^2;
\end{eqnarray*}
these are
consistent with expectations for the $B\rt K\ecp$, where
$\ecp\rt \ks\kpi$.  In addition,
the product branching fraction is comparable in magnitude 
to that for the $\ec$,  also in agreement with expectations
for the $\ecp$.  The observed properties
of this system lead us to conclude that we have observed
the $\ecp$. 

We wish to thank the KEKB accelerator group for the excellent
operation of the KEKB accelerator.
We acknowledge support from the Ministry of Education,
Culture, Sports, Science, and Technology of Japan
and the Japan Society for the Promotion of Science;
the Australian Research Council
and the Australian Department of Industry, Science and Resources;
the National Science Foundation of China under contract No.~10175071;
the Department of Science and Technology of India;
the BK21 program of the Ministry of Education of Korea
and the CHEP SRC program of the Korea Science and Engineering
Foundation;
the Polish State Committee for Scientific Research
under contract No.~2P03B 17017;
the Ministry of Science and Technology of the Russian Federation;
the Ministry of Education, Science and Sport of the Republic of  
Slovenia;
the National Science Council and the Ministry of Education of Taiwan;
and the U.S.\ Department of Energy.

\begin{table}[htb]
\begin{center}
\caption{Results of the fit to the data points in
Fig.~\ref{fig:sliced_results}. Only statistical errors are listed. }
\label{table:fit_results}
\begin{tabular}{l c c c} \hline
Peak       & Events   & Mass (MeV/$c^2$) &
                                 $\Gamma^{\rm tot}$(MeV/$c^2$) \\\hline
$\ec$      & $104 \pm 14$ &  $2979 \pm 2$ &   $ 11 \pm 11  $
\\
$\ecp$     & $39  \pm 11$ &  $3654 \pm 6$ &   $ 15^{+24}_{-15} $
\\\hline
\end{tabular}
\end{center}
\end{table}


\vspace*{-0.5cm}
\begin{references}



\bibitem{Rosner}
For a recent review of this subject see J.L.~Rosner,
Comments on Nucl. \& Part. Phys. {\bf 21}, 369 (1999).

\bibitem{Edwards}
K.W.~Edwards~{\etal} (CLEO Collab.), Phys. Rev. Lett. {\bf 86}, 30
(2001).

\bibitem{Aubert}
B.~Aubert~{\etal} (BaBar Collab.), contribution to
the XXXVII$^{th}$ Rencontres de Moriond on QCD and Hadronic
Interactions, Les Arcs, Savoie, France (March, 2002),
hep-ex/0203040.

\bibitem{ffang}
F.~Fang (Belle Collab.), talk at the Conference
on Flavor Physics and CP Violation, 
May 18, 2002, Philadelphia, PA (unpublished).

\bibitem{PDG}
D.E. Groom {\it et al.} (Particle Data Group),
Eur. Phys. J. {\bf C15}, 1 (2000).

\bibitem{SFT} K.T.~Chao, Y.F.~Gu, and S.F.~Tuan,
Commun. Theor. Phys. {\bf 25}, 471 (1996).

\bibitem{QQ_potential} The range of mass 
splittings quoted in the text is based on an application
of formulae in W.~Buchm\"{u}ller and
S-H.H.~Tye,  Phys. Rev. {\bf D24}, 132 (1981).
Other authors have have given predictions
for the $\ecp$ mass; see, for example, 
G.S.~Bali {\it et al.}, Phys. Rev. {\bf D56}, 2566 (1997);
D.~Ebert {\it et al.}, Phys. Rev. {\bf D62}, 034014 (2000);
E.J.~Eichten and C.~Quigg, Phys. Rev. {\bf D49}, 5845 (1994);
and T.A.~Lahde and D.O.~Riska, hep-ph/0112131, submitted to Nucl.
Phys. {\bf A}.

\bibitem{XTAL} C.~Edwards {\etal}~ (Crystal Ball Collab.), 
Phys. Rev. Lett. {\bf 48}, 70 (1982).

\bibitem{E760} T.A.~Armstrong {\etal}~ (E760 Collab.), 
Phys. Rev. {\bf D52}, 4839 (1995); M.~Masuzawa, Ph.D. Thesis, 
Northwestern Univ. report UMI-94-15774 (1993), unpublished;
M.~Ambrogiani {\it et al.}, Phys. Rev. {\bf D64}, 052003 (2001);
P.~Abreu {\it et al.} (DELPHI), Phys. Lett. {\bf B441}, 479 (1998);
and M.Acciarri {\it et al.} (L3), Phys. Lett. {\bf B461}, 155 (1999).

\bibitem{CC}
Throughout this Letter, whenever a mode is quoted the inclusion
of the charge conjugate mode is implied.

\bibitem{Belle}
A.~Abashian {\it et al.} (Belle Collab.),
Nucl. Instr. and Meth. {\bf A479}, 117 (2002).

\bibitem{KEKB}
 E.~Kikutani ed., KEK Preprint 2001-157 (2001),
 to appear Nucl. Instr. and Meth. A.

\bibitem{SFW}
 The Fox-Wolfram moments were introduced in
 G.C. Fox and S. Wolfram, Phys. Rev. Lett. {\bf 41}, 1581 (1978). 
 The modified moments used in this analysis are described in 
 K. Abe {\it et al.} (Belle Collab.), Phys. Lett. {\bf B 511}, 151 (2001).

\bibitem{MC} Events are generated with the CLEO group's QQ program
(www.lns.cornell.edu/public/CLEO/soft/QQ); the detector
response is simulated using GEANT, R.~Brun~{\etal}, GEANT~3.21,
CERN Report DD/EE/84-1, 1984. 

\bibitem{ARGUS}
H.~Albrecht {\it et al.} (ARGUS Collab.),
Phys. Lett. {\bf B241}, 278 (1990).

\bibitem{Branden} G.~Brandenburg~{\etal} (CLEO Collab.),
Phys. Rev. Lett. {\bf 85}, 3095 (2000).

\bibitem{accept} The acceptance for the $K^+\ks\kpi$ and
$\ks\ks\kpi$ final states 
are 14.0\% and 12.4\%, respectively. (These do not include
reductions due to the $\ks\rt\pipi$ decay branching fraction.)


\end{references}
\end{document}